\documentclass[10pt,conference]{IEEEtran}
\usepackage{cite}
\usepackage{amsmath,amssymb,amsfonts}
\usepackage{algorithmic}
\usepackage{graphicx, float}
\usepackage{textcomp}
\usepackage{xcolor}
\def\BibTeX{{\rm B\kern-.05em{\sc i\kern-.025em b}\kern-.08em
    T\kern-.1667em\lower.7ex\hbox{E}\kern-.125emX,nmlm,x5}}

\usepackage{graphicx}% Include figure files
\usepackage{dcolumn}% Align table columns on decimal point
\usepackage{bm}% bold math
\usepackage{comment}
\usepackage{braket}
\usepackage{booktabs}
\usepackage[export]{adjustbox}

\usepackage[breaklinks=true, colorlinks=true, linkcolor={blue!90!black}, citecolor={blue!90!black}, urlcolor={blue!90!black}]{hyperref}

\newcommand*{\Prob}{\mathsf{Pr}}
\newcommand{\Gate}[1]{\textsc{#1}}

\newcommand{\zgate}{\Gate{z}}

\newcommand{\idgate}{\Gate{i}}

\usepackage{caption}
\usepackage{etoolbox}
\makeatletter
\patchcmd{\@makecaption}
  {\scshape}
  {}
  {}
  {}
\makeatother

\begin{document}
\bstctlcite{IEEEexample:BSTcontrol}

\title{Augmenting QAOA Ansatz with Multiparameter Problem-Independent Layer}

\author{\IEEEauthorblockN{Michelle Chalupnik\IEEEauthorrefmark{1,2}, Hans Melo\IEEEauthorrefmark{2}, Yuri Alexeev\IEEEauthorrefmark{3}, and Alexey Galda\IEEEauthorrefmark{2}}
\IEEEauthorblockA{\IEEEauthorrefmark{1}Department of Physics, Harvard University, Cambridge, MA 02138, USA}
\IEEEauthorblockA{\IEEEauthorrefmark{2}Menten AI, Inc., San Francisco, CA 94111, USA}
\IEEEauthorblockA{\IEEEauthorrefmark{3}Computational Science Division, Argonne National Laboratory, Lemont, IL 60439, USA}
Email: \IEEEauthorrefmark{1}chalupnik@g.harvard.edu, \IEEEauthorrefmark{2}alexey.galda@menten.ai}

\maketitle

\begin{abstract}
The quantum approximate optimization algorithm (QAOA) promises to solve classically intractable computational problems in the area of combinatorial optimization. A growing amount of evidence suggests that the originally proposed form of the QAOA ansatz is not optimal, however. To address this problem, we propose an alternative ansatz, which we call QAOA+, that augments the traditional $p = 1$ QAOA ansatz with an additional multiparameter problem-independent layer. The QAOA+ ansatz allows  obtaining higher approximation ratios than $p = 1$ QAOA while keeping the circuit depth below that of $p = 2$ QAOA, as benchmarked on the MaxCut problem for random regular graphs. We additionally show that the proposed QAOA+ ansatz, while using a larger number of trainable classical parameters than with the standard QAOA, in most cases outperforms alternative multiangle QAOA ansätze.
\end{abstract}

\maketitle

\section{Introduction}
As the noisy intermediate-scale quantum (NISQ) computing era matures, more and more powerful quantum computers become available, making steady progress toward surpassing the computational capabilities of classical computers in solving some important real-world problems. In the field of biochemistry alone, quantum annealers have been used to solve the protein-folding problem \bstctlcite{IEEEexample:BSTcontrol} \cite{Johnson2011, menten_dwave}, and gate-based quantum computers have been used to calculate the ground state energies of molecules \cite{lanyon2010, peruzzo2013, omalley2016, kandala2017, luengo2019}. Despite the ongoing advancements in quantum hardware, however, quantum circuits that can be executed reasonably well by modern quantum computers are still severely limited in the maximum number of qubits, qubit connectivity, and  maximum circuit depth, because of the compounded noise, reducing algorithmic performance well below the best-case theoretical predictions \cite{preskill2018, mitigating_noise, herrman2021,alexeev2021quantum}. 

Hybrid quantum-classical algorithms attack problems through a combination of quantum and classical computational resources and are leading candidates for achieving quantum advantage in near-term devices \cite{sim2019}. Hybrid quantum-classical algorithms range from quantum machine learning techniques \cite{qml} to parameterized quantum circuits, which are quantum circuits with gate parameters that are obtained by using iterative classical optimization routines. Parameterized quantum circuits include quantum variational eigensolvers \cite{mcclean2016, moll2018}, quantum autoencoders \cite{romero2017}, and the quantum approximate optimization algorithm (QAOA) \cite{qaoa}.

QAOA in particular has drawn interest for its potential to show quantum advantage in NISQ devices as well as its numerous practical applications. QAOA solves the combinatorial optimization problem of finding the bitstring corresponding to the lowest energy eigenstate of a Hamiltonian \cite{qaoa}. Performance of QAOA has been improved by exploiting symmetries of graph structures \cite{franca_patron, error_mitigation_galda, graph_structure_herrman, wang2018}, by introducing classical neural networks or other classical methods to assist in parameter optimization \cite{verdon2019, dong2020}, by modifying the circuit ansatz \cite{egger2021, zhu2021, liu2022layer}, and by increasing circuit parameterization, at the expense of increased classical optimization \cite{multiangleqaoa}. Although increasing the number of independent parameters in QAOA circuits can improve performance in some instances, as the number of parameters increases, nonconvex optimization landscapes can hamper parameter optimization \cite{wiersema2020, wang2020, natcom2018mcclean, natcom2021cerezo, shaydulin2019evaluating}. 

In this work we introduce the technique of augmenting the standard QAOA ansatz for the MaxCut problem with a problem-independent layer to improve the performance of $p = 1$ QAOA while keeping the circuit depth below that of $p = 2$ QAOA. The added layer consists of a sequence of two-qubit gates entangling all qubits with the minimum possible number of operations, followed by the second mixing layer of one-qubit gates. We test the performance of our ansatz, which we call QAOA+, on the MaxCut problem for random regular graphs with up 14 nodes, and we show that it yields approximation ratio improvements at the expense of increasing the number of classical optimization parameters. We also compare QAOA+ with another QAOA variant, multiangle QAOA (ma-QAOA)~\cite{multiangleqaoa} and show that for a low number of independent parameters QAOA+ in general results in higher approximation ratios than does ma-QAOA on the MaxCut problem for random regular graphs with up to 14 nodes. 

\section{Background}
In this section we introduce the notation used in this work and briefly review the relevant prior results and approaches for solving the MaxCut problem using QAOA.

\subsection{QAOA}

QAOA is a variational algorithm inspired by the adiabatic evolution principle. In its original formulation \cite{gutmann2020} it can be seen as a Trotterized version of  quantum annealing evolution with a finite number of steps, $p$. QAOA can be used to solve a broad class of general unconstrained combinatorial problems, seeking the solution represented as a bitstring that maximizes the cost function. This is achieved by preparing a parameter-dependent $n$-qubit quantum state using a sequence of alternating operators:
\begin{equation}
    \ket{\vec{\beta}, \vec{\gamma}} := U_B(\beta_p)U_C(\gamma_p)\ldots U_B(\beta_1)U_C(\gamma_1)\ket{s},
\end{equation}
where $U_C(\gamma) = e^{-i\gamma C}$ is the phase operator, $C$ is the problem Hamiltonian, $U_B(\beta) = e^{-i\beta B}$ is the mixing operator, $B = \sum_jX_j$, and $\ket{s}=\ket{+}^{\otimes n}$ is the initial uniform superposition product state, the exited state of $B$ with maximum energy. The evolution is followed by the measurement in the computational basis, with measurement outcome probabilities following the Born rule: 
\begin{equation}
    \Prob(x) = |\braket{x|\vec{\beta},\vec{\gamma}}|^2, \quad x\in \{0,1\}^n.
\end{equation}
The classical parameters $\vec{\beta}, \vec{\gamma}$ are updated in the outer loop to maximize the expected objective value of the quantum evolution measurement outcomes:
\begin{equation}
    \langle C\rangle := \bra{\vec{\beta}, \vec{\gamma}}C\ket{\vec{\beta}, \vec{\gamma}} = \sum_{x\in \{0,1\}^n}\Prob(x)f(x).
\end{equation}
The performance of QAOA relies on the choice of these parameters. The problem of identifying optimal QAOA parameters has generated a large amount of theoretical \cite{wurtz2022counterdiabaticity,boulebnane2021predicting,farhi2019quantum,wurtz2021maxcut} and computational  \cite{lykov2020tensor,medvidovic2021classical,shaydulin2021classical,zhou2020quantum,crooks2018performance,shaydulin2019multistart,shaydulin2021exploiting} work, including such methods as reinforcement learning \cite{reinforcement_learning_wauters, reinforcement_learning_yuri} and reusing preoptimized QAOA parameters from related problem instances \cite{lotshaw2021empirical,transferability,wurtz2021fixed,brandao2018fixed}) to significantly reduce the computational cost of parameter optimization.

\subsection{MaxCut}

MaxCut is an archetypal NP-complete combinatorial problem. Consider an undirected graph $G$ with a set of vertices $V$ and a set of edges $E$. The objective of MaxCut is to find an assignment of vertices into two disjoint subsets such that the number of edges spanning both subsets is maximized. The goal of MaxCut is to maximize the following objective function:
\begin{equation}
    C_{\text{MaxCut}} = \frac{1}{2}\sum_{(j,k)\in E}(\idgate-\zgate_j\zgate_k).
\end{equation}

The performance of QAOA is evaluated with a metric called the approximation ratio (AR). This is defined as the expectation value of the cost Hamiltonian divided by the maximum Hamiltonian energy. 
\begin{align*}
\label{eq:startingeigenstate}
    AR = \langle C \rangle / C_{\text{max}} 
\end{align*}

Gradient-based approaches are often used for QAOA parameter optimization. Unfortunately, the linearly growing number of optimization parameters with the number of QAOA layers, together with the phenomenon of exponentially vanishing gradients, called Barren plateaus \cite{uvarov2020, wang2020, grant2019, mcclean2018barren, cerezo2021cost}, makes this optimization task, in general, an NP-hard nonconvex optimization problem in itself. Increasing the number of QAOA layers also leads to quantum circuits that significantly exceed the capabilities of NISQ hardware, because of limited coherence times and relatively large two-qubit error rates. At the same time, shallow QAOA circuits do not have the capacity to achieve high approximation ratios. It is therefore important to adapt the QAOA approach for the most efficient use on modern quantum processors

\subsection{Alternative QAOA ansätze}

The originally proposed form of the QAOA ansatz \cite{qaoa} has not been proven to be optimal for any particular class of optimization problems. In fact,  a growing amount of research suggests that  alternative QAOA ansätze~\cite{zhu2020adaptive, multiangleqaoa}   outperform the standard QAOA approach, either by requiring shallower quantum circuits to match the performance of standard QAOA or by offering better results for the same circuit depth, or both. In the former case, these improvements can be achieved by developing a tailored problem- and hardware-specific ansatz, as in the case of ADAPT-QAOA \cite{zhu2020adaptive}.
%Several existing alternative QAOA ansätze can achieve lower circuit depth and higher approximation ratios by adding parameters or selecting gates in a problem-specific way. The Adaptive Derivative Assembled Problem Tailored - Quantum Approximate Optimization Algorithm
This approach uses a gradient-based criterion to build a custom QAOA mixer layer from a predefined operator pool, systematically growing the ansatz one layer at a time. This was shown to improve MaxCut approximation ratios for circuit depths $p>4$, while also resulting in a lower number of variational parameters and a lower CNOT gate count.

Another alternative QAOA ansatz, multiangle QAOA \cite{multiangleqaoa} (ma-QAOA), produces improvements in the approximation ratio for the MaxCut problem at the expense of increasing the number of variational parameters without altering the shape of the standard QAOA ansatz. This, however, significantly increases the computational cost of the classical optimization process. Multiangle QAOA independently parameterizes each gate in both the phase separation and mixing layers of the standard QAOA circuit. It has the benefit of requiring lower circuit depths to produce approximation ratios similar to or higher than those achieved with a standard QAOA ansatz. As a result, ma-QAOA offers higher performance than standard QAOA achieves on modern quantum hardware, where increasing circuit depth beyond a certain point leads to a sharp drop in algorithm performance due to decoherence.

\section{QAOA+}

\begin{figure}[ht]
\centering
\includegraphics[width=9cm]{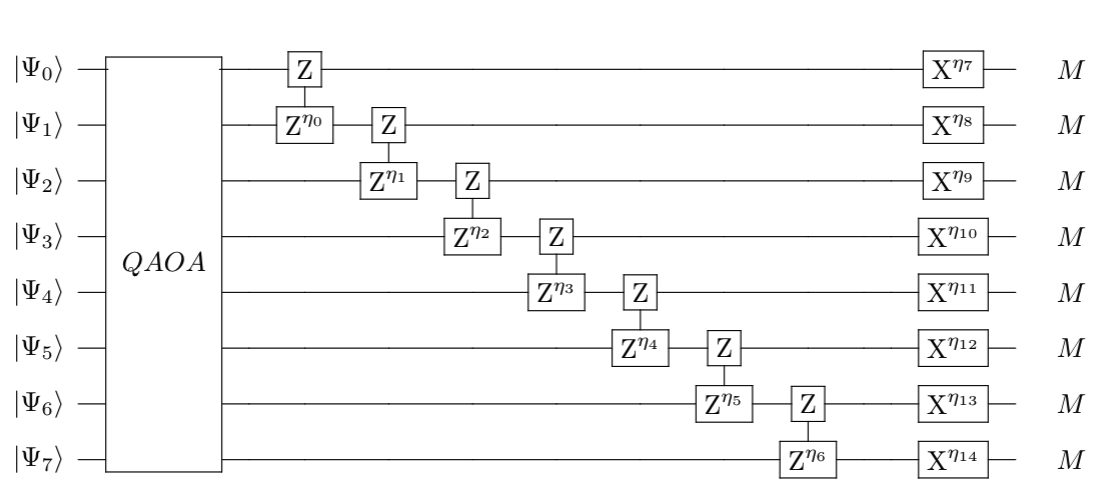}
\caption{Circuit layout for QAOA+ consisting of a standard $p=1$ QAOA layer, followed by an additional problem-independent layer consisting of of $ZZ$ gates, and an additional mixer layer with $X$ gates. All single- and two-qubit gates in the additional layers are independently parameterized.}
\label{fig:schematic}
\end{figure}

We present an alternative QAOA ansatz, called QAOA+, that consists of the standard $p = 1$ QAOA layer followed by an additional problem-independent circuit ansatz (see Fig. \ref{fig:schematic}).
The appended layer has the form of the standard QAOA layer for MaxCut on a $n$-node line graph; in other words,  the phase separation layer sequentially connects all qubits from the first to the last. This problem-independent layer uses a minimum possible number $n - 1$ of two-qubit gates connecting all qubits. Additionally, similarly to ma-QAOA, we allow all variational parameters in the appended layer to be independent. Note that the number of classical parameters in the QAOA+ ansatz is almost always smaller than in ma-QAOA for the same graph.

\begin{table*}[!ht]
\centering
\begin{tabular}{l|lll|lll|lll|lll|}
\cline{2-13}
                                  & \multicolumn{3}{c|}{n = 8}                                      & \multicolumn{3}{c|}{n = 10}                                     & \multicolumn{3}{c|}{n = 12}                                     & \multicolumn{3}{c|}{n=14}                                       \\ \cline{2-13} 
                                  & \multicolumn{1}{l|}{3-reg} & \multicolumn{1}{l|}{4-reg} & 5-reg & \multicolumn{1}{l|}{3-reg} & \multicolumn{1}{l|}{4-reg} & 5-reg & \multicolumn{1}{l|}{3-reg} & \multicolumn{1}{l|}{4-reg} & 5-reg & \multicolumn{1}{l|}{3-reg} & \multicolumn{1}{l|}{4-reg} & 5-reg \\ \hline
\multicolumn{1}{|l|}{QAOA (p=1)}  & \multicolumn{1}{l|}{0.783}     & \multicolumn{1}{l|}{0.785}     & 0.796     & \multicolumn{1}{l|}{0.793}     & \multicolumn{1}{l|}{0.803}     & 0.807     & \multicolumn{1}{l|}{0.768}     & \multicolumn{1}{l|}{0.816}     & 0.814     & \multicolumn{1}{l|}{0.782}     & \multicolumn{1}{l|}{0.812}     & 0.798     \\ \hline
\multicolumn{1}{|l|}{QAOA+ (p=1)} & \multicolumn{1}{l|}{0.838}     & \multicolumn{1}{l|}{0.867}     & 0.866     & \multicolumn{1}{l|}{0.833}     & \multicolumn{1}{l|}{0.844}     & 0.844     & \multicolumn{1}{l|}{0.800}     & \multicolumn{1}{l|}{0.847}     & 0.848     & \multicolumn{1}{l|}{0.815}     & \multicolumn{1}{l|}{0.843}     & 0.827     \\ \hline
\multicolumn{1}{|l|}{QAOA (p=2)}  & \multicolumn{1}{l|}{0.870}     & \multicolumn{1}{l|}{0.884}     & 0.864     & \multicolumn{1}{l|}{0.868}     & \multicolumn{1}{l|}{0.874}     & 0.873     & \multicolumn{1}{l|}{0.840}     & \multicolumn{1}{l|}{0.884}     & 0.882     & \multicolumn{1}{l|}{0.856}     & \multicolumn{1}{l|}{0.883}     & 0.862     \\ \hline
\end{tabular}
\caption{\label{tab:table} Approximation ratios for all 3-, 4-, and 5-regular graphs for $n = 8$, $n = 10$, $n = 12$, and $n = 14$ nodes, for $p = 1$ QAOA, QAOA+, and $p = 2$ QAOA.}
\end{table*}

For dense graphs, in which the number of edges approaches quadratic scaling with the number of nodes, the additional circuit depth associated with the appended linear ansatz layer of QAOA+ is significantly smaller than that of one additional QAOA layer of $p = 2$ QAOA. As a result, QAOA+ offers a favorable compromise between improving the MaxCut approximation ratio, as we show below, and increasing the total circuit depth. This remains true for sparse graphs, although with a lesser effect. Only for MaxCut on linear graphs, when the shape of the standard QAOA ansatz matches that of the additional layer in QAOA+, does the circuit depth advantage of QAOA+ over QAOA disappear. Instead, for linear graphs, the QAOA+ ansatz has the same shape as the standard $p = 2$ QAOA, minus the number of variational parameters.

%\twocolumngrid

\begin{figure}
\includegraphics[width=1.05\linewidth]{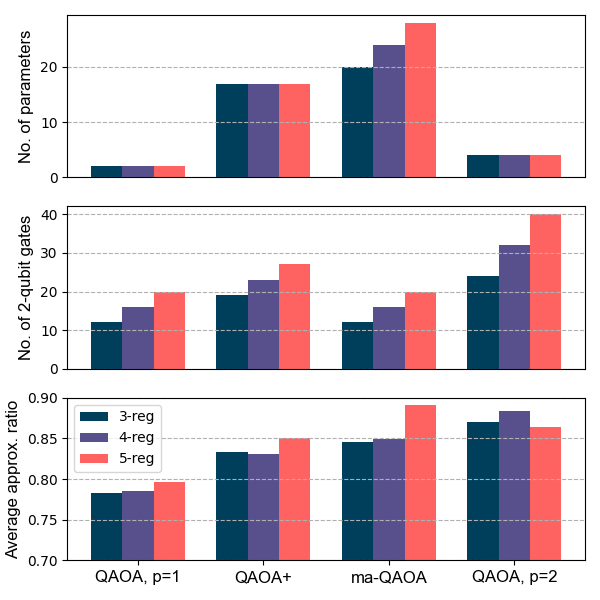}
\caption{Number of parameters, number of two-qubit gates, and averaged approximation ratio for all 3-reg, 4-reg, or 5-reg 8-qubit graphs, plotted for four different QAOA variants: QAOA  ($p=1$), QAOA+ $(p=1)$, multiangle QAOA ($p=1$), and QAOA ($p=2$). }
\label{fig:algcomparison}
\end{figure}

Table \ref{tab:table} compares the mean approximation ratios for MaxCut on non-isomorphic 3-, 4-, and 5-regular 8-, 10-, 12-, and 14-node graphs for $p=1$ QAOA, QAOA+, and $p=2$ QAOA. The results were averaged over 10 non-isomorphic random graphs of each type, whenever available. We used the Broyden--Fletcher--Goldfarb--Shanno (BFGS) algorithm~\cite{press1996numerical} to optimize the QAOA parameters, selecting the best solution from 10 optimization runs with random initial parameters for each graph. QAOA energy expectation values were calculated by using the tensor-network quantum simulator QTensor~\cite{lykov2020tensor,lykov2021performance,lykov2021importance}. The approximation values for the standard $p=1$ and $p=2$ QAOA were obtained from the QAOAKit database~\cite{qaoakit}. Across all considered graph types, the QAOA+ ansatz yields approximation ratio values between those of $p=1$ and $p=2$ QAOA, while also using an intermediate number of two-qubit gates; see Fig.~\ref{fig:algcomparison} for a more detailed comparison of the number of optimized parameters, two-qubit gates, and approximation ratios for 8-node graphs. Since two-qubit gates are the primary source of errors in quantum circuits, the QAOA+ ansatz has the potential of producing better results than the standard $p=2$ QAOA produces when executed on NISQ devices because of the smaller number of two-qubit gates.

\begin{figure}
\includegraphics[width=1.0\linewidth]{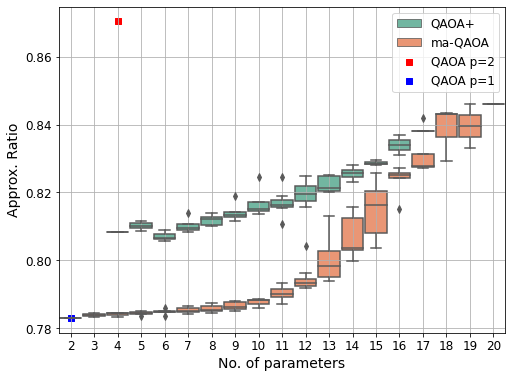}
\caption{Comparison of approximation ratios for ma-QAOA, standard QAOA ($p=1$ and $p=2$), and QAOA+  for all 8-qubit 3-reg graphs. For ma-QAOA and QAOA+, the number of independent parameters are varied to compare algorithm performance at equal parameter number.} 
%QAOAkit \cite{qaoakit} is used as a database of optimized p=1 and p=2 angles to get data points for the standard QAOA.}
\label{fig:pperformance}
\end{figure}

The introduced QAOA+ ansatz not only provides a good compromise between the $p = 1$ and $p = 2$ QAOA in terms of the MaxCut solution quality and the associated QAOA circuit depth, it also offers additional flexibility in the complexity of classical optimization of the QAOA parameters. The additional layer of gates in QAOA+ can be designed to have anywhere between 2 and $2n + 1$ parameters (2 in the standard $p = 1$ QAOA, $n - 1$ in the problem-independent layer of two-qubit gates, and $n$ in the additional mixer layer), where $n$ is the number of qubits. However, the standard QAOA ansatz can also be modified to have more than 2  parameters per single QAOA layer~\cite{multiangleqaoa}. In that work the authors  introduced the ma-QAOA ansatz that allows all parameters in the standard QAOA layer to be different. Because the original ma-QAOA ansatz has no fewer than $2n - 1$ parameters, however, in order to compare the performance of QAOA+ with ma-QAOA, we have modified both ansätze to have an adjustable number of parameters, ranging from 4 to $2n + 1$ for QAOA+ and from 2 to the maximum problem-dependent number of parameters in ma-QAOA. Moreover, because the number of parameters can be unevenly distributed between the two-qubit and mixer layers, for both QAOA+ and ma-QAOA we have averaged the approximation ratios over QAOA instances with all possible distributions of the optimization parameters between. For each instance, we have optimized all five possible non-isomorphic 8-node 3-regular graphs using the BFGS algorithm with 10 restarts. The results are shown in Fig.~\ref{fig:pperformance}.

The number of independent optimization parameters determines the complexity of classical optimization of QAOA. We have observed that QAOA+ always gives higher approximation ratios than ma-QAOA does at equal parameter numbers, despite the fact that with the full 20 parameters ma-QAOA outperforms the maximally parameterized 17-parameter QAOA+. The gap in performance between QAOA+ and ma-QAOA increases as the number of optimization parameters is reduced. In fact, the 4-parameter QAOA+ performs better on average than 14-parameter ma-QAOA does, while still doing worse than $p = 2$ QAOA with a larger circuit depth; see Fig.~\ref{fig:pperformance}.

\begin{figure}
\centering
\includegraphics[width=0.9\linewidth]{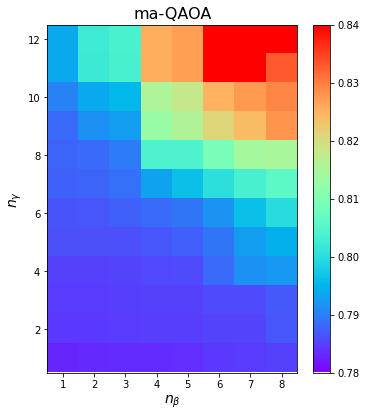}
\caption{Average approximation ratio for all non-isomorphic 8-node 3-regular graphs, using ma-QAOA with varied number of parameters. The total number of independent circuit parameters equals $n_{\gamma}$ + $n_{\beta}$.}
\label{fig:maqaoa_per_param}
\end{figure}

We have not observed any significant difference in performance between ma-QAOA ansätze with the same number of parameters but distributed differently between the cost and mixing layers; see Fig.~\ref{fig:maqaoa_per_param}. While 
%it is inconclusive from Fig.~\ref{fig:maqaoa_per_param} 
the figure does not make clear whether the QAOA+ ansatz favors circuit parameterizations with any particular distribution of the number of parameters between the auxiliary two-qubit layer and the second mixing layer, one can clearly see that QAOA+ yields higher approximation ratios than ma-QAOA+ does for all possible combinations of parameters.

\begin{figure}
\centering
\includegraphics[width=0.9\linewidth]{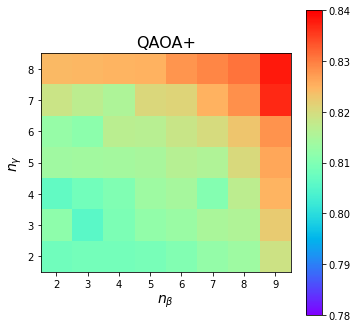}
\caption{Average approximation ratio for all non-isomorphic 8-node 3-regular graphs, using QAOA+ with varied number of parameters. Here, $\beta$ and $\gamma$ refer to the number of optimized parameters in the ansatz layer. The total number of independent circuit parameters equals $n_{\gamma}$ + $n_{\beta}$.}
\label{fig:qaoaplus_per_param}
\end{figure}

\begin{figure}
\includegraphics[width=0.95\linewidth]{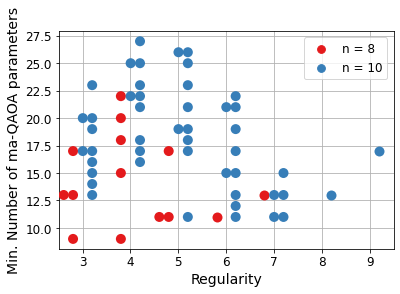}
\caption{Minimum number of parameters required for ma-QAOA to match the performance of QAOA+(4) for $d$-regular 8-node and 10-node graphs.}
\label{fig:8_and_10_qubit_thresholds}
\end{figure}

While Fig.~\ref{fig:algcomparison} presents a compelling case that QAOA+ with 4 optimization parameters, which we will call QAOA+(4) for brevity, gives higher approximation ratios than does ma-QAOA  with significantly more parameters, these results are limited to the MaxCut problem on 3-regular 8-node graphs. To establish this fact for larger MaxCut instances, we  calculated the approximation ratio for the QAOA+(4) ansatz for all types of 8- and 10-node random regular graphs, averaged over 10 graphs of each type and using the BFGS optimization with 10 restarts. We then perform the same averaging procedure for the ma-QAOA ansatz on the same graphs with an increasing number of optimization parameters from 2 until the approximation ratio of QAOA+(4) is exceeded, recording that minimum number of parameters in ma-QAOA necessary to outperform QAOA+(4). The results are shown in Fig. \ref{fig:8_and_10_qubit_thresholds}. Unlike in Fig. \ref{fig:pperformance}, only one parameter distribution between the cost and mixing layers in the ma-QAOA ansatz is considered: for every even number of parameters, they are split equally between the two layers, while for an odd total number of parameters the cost layer had one more parameter. The thresholds after which ma-QAOA yields higher approximation ratios are averaged over up to 10 random regular graphs of each type. At parameter numbers below this threshold, QAOA+(4) outperforms ma-QAOA, while above the threshold, ma-QAOA starts to produce higher approximation ratios, at the expense of a significantly large-dimensional search space of QAOA parameters. As the size of regular graphs is increased from 8 to 10 nodes, the threshold values rise, indicating that the performance of our QAOA+ ansatz scales favorably with problem size, compared with ma-QAOA.

\section{Discussion}

Increasing the circuit depth by increasing the number of QAOA layers is necessary in order to improve the approximation ratio achievable with the standard QAOA approach. When executing on NISQ hardware, however, the increased circuit depth comes with a severe penalty of decreasing performance due to decoherence. Alternative QAOA ans\"{a}tze aim to modify the QAOA approach to increase MaxCut approximation ratios at reduced costs in terms of circuit depth. In this work we have introduced the QAOA+ as an alternative ansatz that improves approximation ratios compared with the standard $p=1$ QAOA by adding an auxiliary layer of gates that is smaller than a full additional QAOA layer. We have shown that QAOA+ provides a viable alternative for an intermediate ansatz with approximation ratios and circuit depth between those of $p=1$ and $p=2$ QAOA. 

Because the introduced QAOA+ ansatz allows for a large number of classical optimization parameters, as a baseline we compared QAOA+ with ma-QAOA, another alternative QAOA ansatz with low circuit depth and high number of parameters. While ma-QAOA generally yields higher approximation ratios than does QAOA+ when all optimization parameters are utilized, we have demonstrated that when the number of parameters in QAOA+ and ma-QAOA is the same,  QAOA+ significantly outperforms ma-QAOA, with the performance gap growing with the problem size. In particular, we  considered QAOA+ with 4 parameters and demonstrated that ma-QAOA requires significantly more optimization parameters to match the performance of QAOA+, which comes at large classical computation overhead. Therefore, in general, QAOA+ requires fewer parameters than does ma-QAOA for the same level of performance, while having a larger circuit depth. When minimizing the objective function for many parameters is difficult, QAOA+ will be a valuable option. The reduction in circuit depth from using QAOA+ compared with the standard $p=2$ QAOA is particularly noticeable for MaxCut instances on dense graphs.

This work does not present an exhaustive study on all possible alternative QAOA ansätze. We have  considered only a single shape of the additional problem-independent layer of two-qubits gates, defined by the minimum possible number of gates that entangle all qubits. Other shapes of the QAOA+ ansatz will be considered elsewhere, including more hardware-efficient ansätze that respect the possible connectivity constraints of quantum processors. One can also consider such modification as creating a hybrid ma-QAOA/QAOA+ ansatz consisting of both the multiparameter ma-QAOA and the additional multiparameter problem-independent QAOA+ ansatz layers, in order to shift the improvements in the MaxCut approximation ratio even more toward the classical computation instead of increasing the QAOA circuit depth. The study of the necessary computational resources required for optimization of the increased number of optimization parameters in the proposed QAOA+ ansatz,  as well as the associated challenged of Barren plateaus, was also beyond the scope of this work. Performance comparisons presented here were done under the assumption of equal computational limitation on optimization across all optimization instances regardless of the number of parameters. Additionally, we  considered only the $p=1$ case, and performance can be analyzed at larger circuit depths. Moreover, because this work was motivated by potential performance improvements during execution on NISQ hardware, noisy simulations and comparisons on modern quantum processors will need to be performed to validate the proposed approach.

When high-fidelity quantum computing hardware capable of executing large-scale QAOA instances with $p \gg 1$ becomes available, the true value of alternative QAOA ansätze will be revealed. The optimal balance between the computational complexity of parameter optimization and the acceptable depth of quantum circuits will be determined through extensive experimentation. Until then, it is crucial to continue improving the capacity of the QAOA approach to solve NP-hard combinatorial optimization problems by developing novel ansätze. We are hopeful that the approach introduced in our paper will help advance the standard QAOA of today toward the next-generation QAOA of tomorrow capable of unlocking practical quantum advantage.

\section*{Acknowledgments}
We thank Gavin Crooks, Vikram Mulligan, and Ian MacCormack for helpful discussions and feedback on the manuscript. M.C. was supported by funding provided during an internship at Menten AI and partially supported by the Department of Defense (DoD) through the National Defense Science and Engineering Graduate (NDSEG) Fellowship Program, and acknowledges support from her PhD advisor Marko Lon\v{c}ar. This material is partially based upon work supported by the Defense Advanced Research Projects Agency (DARPA) under Contract No. HR001120C0068. Y.A.’s work at Argonne National Laboratory was supported by the U.S. Department of Energy, Office of Science, under contract DE-AC02-06CH11357.

%This material was based upon work supported by the U.S. Department of Energy, Office of Science, under contract number DE-AC02-06CH11357.  

\medskip

\bibliographystyle{IEEEtran}
\bibliography{sources.bib}

%GAIL - need to add govt license (removed when the paper is accepted):
\vfill
\framebox{\parbox{.90\linewidth}{\scriptsize The submitted manuscript has been created by UChicago Argonne, LLC, Operator of Argonne National Laboratory (``Argonne''). Argonne, a U.S.\ Department of Energy Office of Science laboratory, is operated under Contract No.\ DE-AC02-06CH11357.  The U.S.\ Government retains for itself, and others acting on its behalf, a paid-up nonexclusive, irrevocable worldwide license in said article to reproduce, prepare derivative works, distribute copies to the public, and perform publicly and display publicly, by or on behalf of the Government.  The Department of Energy will provide public access to these results of federally sponsored research in accordance with the DOE Public Access Plan \url{http://energy.gov/downloads/doe-public-access-plan}.}}
\clearpage

\end{document}